# Quasiparticle and Optical Spectroscopy of Organic Semiconductors Pentacene and PTCDA from First Principles

Sahar Sharifzadeh,[a] Ariel Biller,[b] Leeor Kronik,[b*] and Jeffrey B. Neaton[a*]

[a] Molecular Foundry, Lawrence Berkeley National Laboratory, Berkeley, CA 94720
[b] Department of Materials and Interfaces, Weizmann Institute of Science, Rehovot, Israel

**Corresponding authors**:

Jeffrey B. Neaton, jbneaton@lbl.gov, phone: (510) 486-4527, fax: (510) 486-7424;

Leeor Kronik, leeor.kronik@weizmann.ac.il, phone: +972-8-934-4993, fax: +972-8-934-4138

# Abstract

The broad use of organic semiconductors for optoelectronic applications relies on quantitative understanding and control of their spectroscopic properties. Of paramount importance are the transport gap – the difference between ionization potential and electron affinity – and the exciton binding energy – inferred from the difference between the transport and optical absorption gaps. Transport gaps are commonly established via photoemission and inverse photoemission spectroscopy (PES/IPES). However, PES and IPES are surface-sensitive, average over a dynamic lattice, and are subject to extrinsic effects, leading to significant uncertainty in gaps. Here, we use density functional theory and many-body perturbation theory to calculate the spectroscopic properties of two prototypical organic semiconductors, pentacene and 3,4,9,10-perylene tetracarboxylic dianhydride (PTCDA), quantitatively comparing with measured PES, IPES, and optical absorption spectra. For bulk pentacene and PTCDA, the computed transport gaps are 2.4 and 3.0 eV, and optical gaps are 1.7 and 2.1 eV, respectively. Computed bulk quasiparticle spectra are in excellent agreement with surface-sensitive photoemission measurements over several eV only if the measured gap is reduced by 0.6 eV for pentacene, and 0.6-0.9 eV for PTCDA. We attribute this redshift to several physical effects, including incomplete charge screening at the surface, static and dynamical disorder, and experimental resolution. Optical gaps are in excellent agreement with experiment, with solid-state exciton binding energies of ~0.5 eV for both systems; for pentacene, the exciton is delocalized over several molecules and exhibits significant charge transfer character. Our parameter-free calculations provide new interpretation of spectroscopic properties of organic semiconductors critical to optoelectronics.

## I. Introduction

Organic semiconductors are highly promising for a variety of electronics, photovoltaics, and spintronics applications.[1-5] They possess many potential advantages over their inorganic counterparts – for example, cheap processing, abundance, low-power consumption, flexibility, and tunability of electronic structure. Pentacene and 3,4,9,10-perylene tetracarboxylic dianhydride (PTCDA), shown in Fig. 1, are two well-studied organic semiconductors that form highly ordered crystalline films.[6] While surface-sensitive photoelectron spectroscopies have been used extensively to characterize their electronic structure,[7-13] the physical interpretation of these measurements has been actively debated:[10-12] for example, reported transport gaps and, by further comparison to optical spectroscopy, exciton binding energies, can differ by up to 1.0 eV. Given the importance of these fundamental quantities for optoelectronic and photovoltaic applications, quantitative insight from electronic structure calculations for these two widely-studied materials would be crucial for understanding these and related spectroscopic quantities.

Density functional theory (DFT) is a method of choice for calculating the electronic structure of extended systems.[14, 15] However, it is well known that DFT, within standard local and semilocal approximations to the exchange-correlation potential, does not quantitatively describe the spectroscopic properties of molecular solids, in particular their transport and optical gaps.[16] First-principles many-body perturbation theory (MBPT),[17, 18] typically based on DFT calculations as a starting point, is a state-of-the-art excited state approach that has been shown to produce transport gaps, band structures, and optical absorption spectra for inorganic solids that are in excellent agreement with experiment,[17-22] and has shown promise for organic solids and gas-phase molecules (see for example Refs.[23-26] for solids and [27-33] for molecules). While other excited-state approaches have been applied to the electronic structure of pentacene and PTCDA,[25, 27, 32, 34-45] there has been no adjustable parameter-free, rigorous comparison with experimental photoemission spectra in the solid-state, and thus the magnitudes and origins of the transport gap and exciton binding energies in these "simple" organic semiconductors remain an open question.

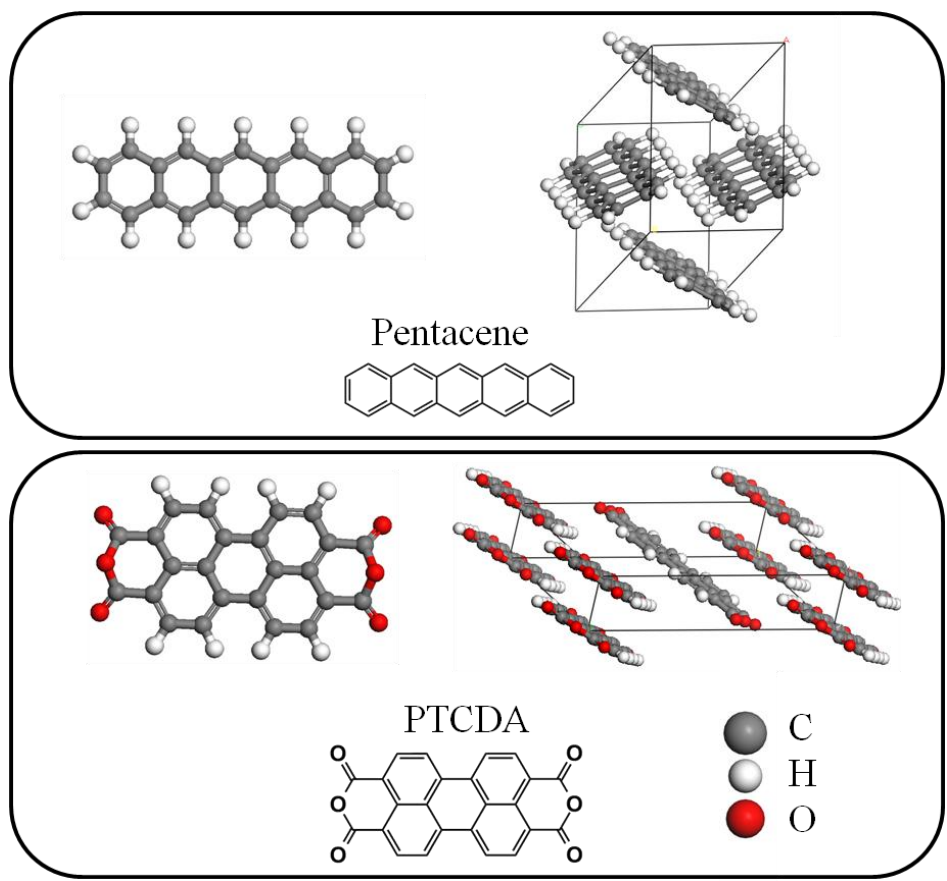

**Figure 1:** Pentacene and PTCDA molecule and bulk crystal.

Here, we use MBPT to compute and quantitatively understand the detailed electronic structure of pentacene and PTCDA, and their transport and optical gaps. We find that, for both the gas and bulk crystalline phases, calculated fundamental and optical gaps are in good agreement with pertinent experiments, allowing us to resolve the uncertainty in the magnitude of the exciton binding energy for the crystal. Moreover, our computed densities of states are in excellent agreement with photoemission experiments – over several eV on either side of the gap – but only if the photoemission spectrum is shifted rigidly such that the gap between occupied and unoccupied spectra is reduced by 0.6-0.9 eV. This shift of the photoemission spectrum, which we attribute to the sum of several different physical factors, sheds light on the relation between the surface spectroscopy and bulk quantities, thus resolving long-standing experimental questions about the spectroscopic properties of organic semiconductors critical to optoelectronics.

## II. Theoretical Approach

### A. Formalism

The formal basis for our MBPT approach is summarized in Refs.[17] and [18]. Briefly, we first determine the quasiparticle (QP) energies and wavefunctions, $\varepsilon_i$ and $\phi_i(r)$, respectively, associated with addition or removal of a charge from the system, by seeking the solution of $H_0\phi_i(r) + \int \Sigma(r,r';\omega=\varepsilon_i)\phi_i(r')d^3r' = \varepsilon_i\phi_i(r)$, where $H_0$ is a Hamiltonian consisting of the non-interacting electron kinetic energy, electron-ion interactions, and the mean-field (Hartree) electron-electron repulsion. The non-local self-energy operator, $\Sigma$, which accounts for electron-electron interactions beyond mean-field, is approximated to first order, within the GW approximation, as the convolution of the screened Coulomb interaction,

$$W(r,r';\omega) = \frac{1}{\Omega}\int \varepsilon^{-1}(r,r'';\omega)V^c(r''-r')d^3r'',$$

and the one-particle electronic Green function, $G(r,r';\omega)$.[46] Here, $V^c$ is the Coulomb potential and $\varepsilon$ the dielectric function.

In the following, we adopt a standard approach[17, 47] in which the QP wavefunctions are approximated as those obtained from DFT. $G(r,r';\omega)$ is approximated as

$$G(r,r';\omega) = \sum_{n,k} \frac{\phi_{nk}(r)\phi^*_{nk}(r')}{\omega - E^{DFT}_{nk}},$$

where $\phi_{nk}(r)$ and $E^{DFT}_{nk}$ are the DFT eigenvalues and eigenvectors respectively, and $n$ and k are the DFT band and k-point index. Following previous work, the static dielectric matrix is calculated within the random phase approximation (RPA) and is extended to finite frequency, $\omega$, with a generalized plasmon pole (GPP) model.[17] We confirm the validity of the GPP for the bulk crystals by comparison with full frequency (FF) calculations.[48-50] From the self-energy, the QP energies are computed as a first-order correction to DFT energies.

So far, we have defined the perturbative approach, known as $G_0W_0$, which has no self-consistency on the eigenvalues and eigenvectors that enter into G and W. This approximation should work best if the KS eigensystems are close to the QP values. To test the approximation, we performed additional calculations, where a $G_0W_0$ QP spectrum is used as the starting eigenvalues spectrum to construct G and W for a subsequent GW calculation; this we label as $G_1W_1$. Additionally, to understand the influence of the starting DFT functional, we consider both the standard generalized gradient approximation of Perdew, Burke, and Ernzerhof (GGA-PBE)[51] and the short-range hybrid functional of Heyd, Scuseria and Ernzerhof (HSE)[52-54] as a starting point for GW. Unless otherwise stated, we perform $G_0W_0$ calculations with a periodic, planewave DFT-PBE starting point within GPP-GW, which is labeled as $G_0W_0$(PBE) for convenience.

Because periodic boundary conditions are imposed, isolated systems are placed in a large supercell, with lattice vectors along the non-periodic directions that are twice the size necessary to contain 99.9% of the charge density; and, in order to avoid spurious interactions between periodic images, the Coulomb potential is truncated at half of the unit cell length.

Given the static inverse dielectric function computed within the RPA and QP energies, neutral excitation energies and spectra are found via the solution of the Bethe-Salpeter equation (BSE). We use an approximate form of the BSE developed within an *ab initio* framework by Rohlfing and Louie,[18] which involves solving the equation

$$(E_{ck}^{QP} - E_{vk}^{QP})A_{vck}^S + \sum_{v'c'k'}\langle vck|K^{eh}|v'c'k'\rangle A_{v'c'k'}^S = \Omega^S A_{vck}^S,$$

where $K^{eh}$ is the electron-hole interaction kernel, and the excited state, *S*, is a sum of products of valence (occupied), *v*, and conduction (unoccupied), *c*, DFT eigenstates, which is evaluated within the Tamm-Dancoff approximation, as

$$|S\rangle = \sum_{vck} A_{vck}^S |vck\rangle. \qquad 1$$

$A_{vck}^S$ are the coupling coefficients, and $\Omega^S$ is the eigenvalue of *S*.

### B. Computational Details

The molecular structure of pentacene and PTCDA is shown in Fig. 1. Both molecules have a $D_{2h}$ point-group symmetry in the gas phase and crystallize with 2 molecules per primitive cell, with pentacene having triclinic (P-1) and PTCDA monoclinic (P2$_1$/c) space group symmetry. Both solids are known to be polymorphic;[55, 56] we restrict our studies to the S-phase of pentacene[57] and α-phase of PTCDA.[56] The lattice vectors are kept fixed to experiment since PBE does not include the van der Waals interactions necessary for describing intermolecular spacing in these weakly-bound crystals.[58-60] The molecular geometry is optimized within GGA-PBE using the Vienna *ab initio* simulation package (VASP),[61, 62] and default projector augmented wave (PAW) potentials.[63] Here 1, 4, and 6 electrons are considered explicitly as valence electrons for H, C, and O, respectively. We use a Γ-centered k-point mesh of 2x2x1 (2x3x2) and a wavefunction cutoff of 400 eV (500 eV) for pentacene (PTCDA). These parameters are sufficient to converge the total energy to 1meV/atom.

GW/BSE calculations are performed with the BerkeleyGW package,[64] with convergence parameters increased such that near-gap states are converged to 0.01 eV within the approximations described here. Starting DFT-PBE eigenvectors and eigenvalues were taken from the Quantum Espresso DFT package,[65] which is compatible with the BerkeleyGW implementation. Here, Troullier-Martins norm-conserving pseudopotentials[66] are employed to represent the core electrons and nuclei, and the number of valence electrons for each atom is the same as those used in the molecular structure studies. Default core radii are used (1.5 au for C, 1.3 au for H, and 1.4 au for O), and pseudopotentials are tested against those with smaller core radii (1.3 au for C, 1.0 au for H, and 1.3 au for O), which yield a change of 0.04 eV or less in the orbital eigenvalues at the DFT level. For the softer pseudopotentials, a wavefunction kinetic energy cutoff of 680 (816) eV is used for pentacene (PTCDA) while for the harder pseudopotential, the cutoff is 1090 (1225) eV. The GW dielectric matrix cutoff is 136 eV.

The calculation of Σ involves a sum over the full subspace of unoccupied states which is truncated in practice.[47] We converge the sum by increasing the number of empty states until the gap changes by less than 0.02 eV. 3060, 3263, 466, and 480 unoccupied states are used for the pentacene molecule, PTCDA molecule, pentacene crystal, and PTCDA crystal, respectively. GW states that are located within 4 eV of the gap were included in the BSE summation, with a k-point mesh of 4x4x2 for pentacene and 2x2x2 for PTCDA describing the excited state. For pentacene, the k-point mesh is larger than necessary for DFT in order to properly describe the delocalization of the excited state.

Our tests of full frequency GW (FF-GW) calculations are performed using the VASP code which utilizes a spectral representation of the dielectric matrix.[48-50] FF-GW convergence parameters are the same as those within the GPP as noted above, except that the dielectric matrix cutoff is 100 eV and the number of frequency points along the real axis are 64. The differences between QP eigenvalues predicted by GPP and FF-GW are minor for states closest to the gap and increase for states farther from the gap, as expected. The eigenvalues for the highest occupied and lowest unoccupied energy bands, averaged over k, change by less than 0.02 eV (0.03 eV) for pentacene (PTCDA) and states up to ±4 eV away from the gap change by less than 0.2 eV (0.35 eV) for pentacene (PTCDA). Given such good agreement between GPP-GW and FF-GW eigenvalues, we report only GPP results in the text, as calculated with the BerkeleyGW code,[64] with the exception of $G_1W_1$ and $G_0W_0$ with a DFT-HSE starting point for solids, which are performed with FF-GW, for computational convenience.

## III. Results and Discussion

In the following, in Section III.A, we first benchmark and validate our theoretical approach for predicting the nature and energy of low-lying vertical excitations within the organic molecules and crystals. Subsequently, in Section III.B, by detailed analysis of the predicted bandstructure and density of states, we discuss the efficacy of photoemission spectroscopy in providing bulk transport gaps of organic semiconductors.

### A. Fundamental and optical gaps

We first consider the vertical -- i.e., no geometry relaxation upon excitation -- addition/removal gap of pentacene and PTCDA (Table 1), and compare the predicted gaps with experiment. The highest occupied molecular orbital (HOMO) and lowest unoccupied molecular orbital (LUMO) are of $\pi$-symmetry, as expected for such conjugated molecules. Overall, our calculated fundamental gap for the pentacene molecule compares well with the experimental addition/removal gap, defined here as the difference between the ionization potential (IP) obtained from photoemission spectroscopy[67] and the electron affinity (EA) estimated from the free attachment energy measured by electron-transfer equilibria.[68] Although it is not clear that the latter measurement results in vertical excitations, we expect adiabatic effects to be quite small for the rigid pentacene and PTCDA molecules: within the $\Delta$SCF[69] approach, the IP and EA change by less than 0.1 eV when geometry relaxation in the excited state is considered. $G_0W_0$(PBE) corrections are a drastic improvement on standard DFT-PBE, increasing the gap of 1.1 eV to 4.5 eV; yet they still underestimate the experimental gap for pentacene by 0.7 eV. Upon an update of the eigenvalue spectrum and a subsequent $G_1W_1$ calculation, the difference with experiment is further reduced, with the pentacene gap opening to 4.9 eV (within 0.3 eV of experiment). For PTCDA, the experimental electron affinity (EA) is unavailable and so the accuracy of our $G_0W_0$ gap of 4.7 eV is not known. Nevertheless, the PTCDA gap also opens up with $G_1W_1$ to 5.1 eV, indicating that the introduction of self-consistency at the level of eigenvalues has an effect on the predicted electronic structure.

A residual underestimation of molecular fundamental gaps by $G_0W_0$ has been observed for a variety of molecules,[27, 28, 32, 33] including gas-phase pentacene and PTCDA, with fundamental gaps in good agreement with the results given here.[27, 32] It was suggested that a polarizability matrix, built from a DFT calculation where the Kohn-Sham HOMO-LUMO gap drastically underestimates the fundamental gap (by a factor of 5 for PBE in this case), leads to over-screening and an underestimated GW result for the gap relative to experiment.[32] This is consistent with the fact that $G_1W_1$ leads to better agreement with experiment for pentacene. We also note that the computed GW gaps are consistent with other high-level theoretical predictions. For example, the ΔSCF approach, with which IP and EA are computed as total energy differences, leads to gaps that range from 4.4-4.8 eV for pentacene and 4.7-5.0 eV for PTCDA (the range of numbers is due to the differing starting DFT functional--local, semi-local, and hybrid). Additionally, the values agree well with those predicted by optimally-tuned range-separated hybrid functional DFT calculations,[70, 71] which yield accurate fundamental gaps for finite-sized systems.[70]

In the bulk crystalline state, the highest occupied and lowest unoccupied states of pentacene and PTCDA retain their π-character, as expected from the weak intermolecular coupling. The crystals' vertical fundamental gaps (i.e. "transport" gaps), taken as the difference between the top of the valence band and the bottom of the conduction band, are 2.2 eV for pentacene (consistent with prior work[25]) and 2.7 eV for PTCDA, about half of the value for the isolated molecule in both cases. Though $G_0W_0$ somewhat underestimates the gap of the isolated molecule, likely due to over-screening, such errors are much smaller for the bulk crystal, where the DFT energy spectrum is much closer to the addition/removal or QP spectrum. For both $G_1W_1$(PBE) and $G_0W_0$(HSE), the static dielectric constant is slightly decreased, with respect to $G_0W_0$(PBE) (from 3.6 to 3.2 for pentacene and 4.0 to 3.3 for PTCDA), and the band gap is slightly increased, by 0.2 eV for pentacene and 0.3 eV for PTCDA.

**Table 1:** PBE starting point based calculations for the fundamental and optical gaps of pentacene and PTCDA in the gas and crystalline phases, with experimental values in parentheses.

| (eV) | Molecule | Crystal | Molecule | Crystal |
|---|---|---|---|---|
| | **Pentacene** | | **PTCDA** | |
| PBE gap | 1.1 | 0.75 | 1.5 | 1.3 |
| $G_0W_0$(PBE) fundamental gap | 4.5 (5.2)[67,68] | 2.2 (2.2)[97] | 4.7 | 2.7 (2.5-2.8)[11] |
| Polarization model fundamental gap | ---- | 2.1 | ---- | 2.3 |
| $G_0W_0$(PBE)/BSE optical gap | 2.2 (2.3)[105] | 1.7 (1.8-1.85)[106-108] | 2.6 (2.6)[109] | 2.1(2.2-2.25)[110,111] |
| Exciton binding energy | 2.3 | 0.5 | 2.1 | 0.6 |
| Polarization model exciton binding energy | ---- | 0.6 | ---- | 0.65 |

The role of static polarization on QP excitations in the molecular crystals is significant, and can be used to rationalize the difference in the QP gaps between the gas phase and the solid state. The reduction in the QP gap upon crystallization can be explained primarily by the dielectric screening of the bulk which decreases the energy to add or remove a charge from a molecule. This non-local correlation effect is not captured by DFT with common functionals, as shown previously.[72] We estimate the polarization energy by a simple electrostatic model, with the organic crystal represented as a linear dielectric and the charged molecule as a hollow sphere of radius R, with a point charge placed at its center. The gap of the molecule within the solid is reduced by twice the polarization energy, P, which is $P = -e^2(\varepsilon-1)/(2R\varepsilon)$.[10, 73] Here, e is the magnitude of the electronic charge and P is in atomic units. We obtain R from the volume per molecule in the primitive cell, ($V_{cell}/2$), as $R = [3V_{cell}/8\pi]^{1/3}$, and $\varepsilon$ as the static dielectric constant which we calculate within the RPA. Within $G_0W_0$(PBE), our estimates of the solid-state transport gap obtained by direct computation and by reducing the gas-phase gap by 2P agree within 0.3 eV (see Table 1). The same statement applies to the $G_1W_1$(PBE) results. Moreover, for both molecules, the value of P within this model is ~1.2 eV, within 0.2 eV of that calculated with a charge redistribution model of polarization,[74-76] and with polarizable continuum model studies of pentacene.[77]

Within $G_0W_0$/BSE, the lowest-energy neutral excitation for both gas-phase and bulk crystalline molecules is of $\pi$-$\pi$* character. In Fig. 2, we plot the electronic component of the electron-hole wavefunction (Eq.1), with the hole explicitly placed slightly above a C atom of one of the molecules in the unit cell, a position of high hole probability,[78] in order to provide a representation of the shape and extent of the exciton. Although the nature of the excited state in the crystal shows a similar character to the gas-phase (insets in Fig. 2), the excited state can delocalize to neighboring molecules. Consistent with previous calculations,[25, 39] the exciton in pentacene is delocalized over a few molecules within a plane. Furthermore, the absence of a significant electron density around the hole site of the exciton wavefunction implies a partial charge transfer character upon excitation. (The additional impact of lattice relaxation and its possible effects on the exciton character is relegated to future work.) For excitons in pentacene, such a character has been proposed,[79] but recently questioned,[80] based on electric field modulated absorption studies. For PTCDA, the first excitation is more localized, with significant electron density at the hole location, and also on the nearest neighbor molecules in the π-stacked direction. A qualitatively similar first excited state in the PTCDA crystal has been reported previously.[81]

The lowest vertical excitation energy provides an estimate of the optical gap as extracted from optical ellipsometry and absorption. As shown in Table 1, the predicted and measured gaps agree to 0.15 eV for the molecules and phases considered. We also note that for the organic crystals, the two lowest energy excitations are nearly degenerate (within 0.1 eV), consistent with two nearly equivalent molecules in the unit cell. Here, the QP wavefunction and energies that make up the two-particle wavefunction of Eq. 1 are taken from $G_0W_0$(PBE), which, for technical reasons, is a more straightforward starting point. Although the addition/removal gaps are affected by up to a few tenths of eV by the GW starting point due to over-screening, we expect that optical excitation energies will be much less affected because screening is less significant for a neutral excitation, as evident (for example) by the small difference in the optical gap of the gas and crystal phases.

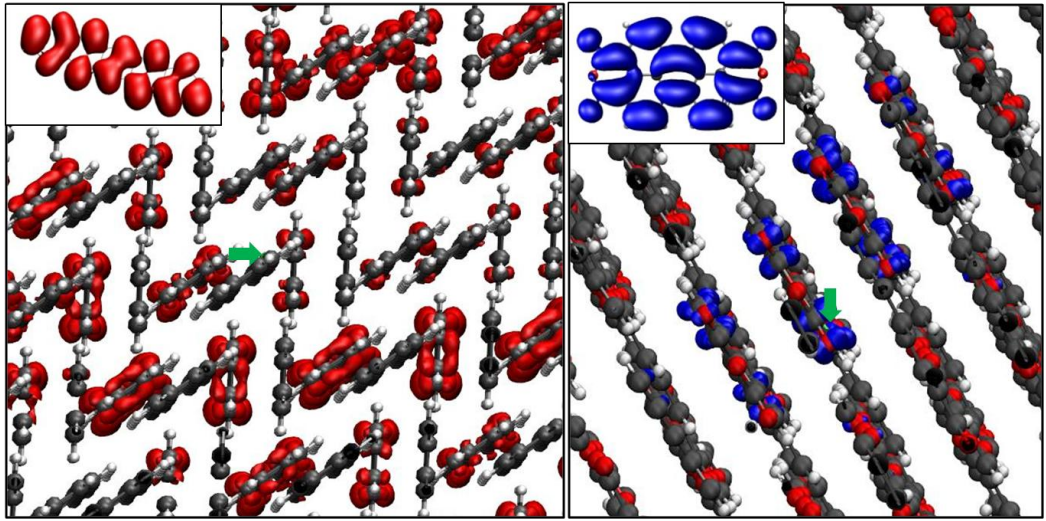

**Figure 2**: The electronic component of the excited state wavefunction (Eq. 1) for pentacene (red) and PTCDA (blue). The hole located in a location of high probability, less than 1 Å above a C atom in the molecular plane, as indicated by a green arrow. Insets show the same for the gas-phase molecule. For the crystal, an isosurface value which contains 30% of the charge density is taken, while for the isolated molecules, the isosurface value is taken to be 1% of the maximum.

The exciton binding energy, defined as the difference between the transport and optical gaps, is determined from the $G_0W_0$(PBE) calculations to be 0.5 eV for pentacene and 0.6 eV for PTCDA. These values are in good agreement with those obtained from DFT/time-dependent DFT studies of solvated pentacene (0.5 eV) and PTCDA (0.6 eV).[44] However, there is a discrepancy between our calculated exciton binding energy for pentacene and the 0.1 eV predicted based on an implementation of BSE that relies on empirical fit of the QP eigenvalues,[35] a difference that may be ascribed to their empirical fit, and possibly also to slight differences in the DFT geometries.

The value of the exciton binding energy, being the difference between the gap of charged excitations and the gap of neutral excitations, is again subject to polarization effects. For both systems, we find it to be consistent with a simple electrostatic model of bulk screening of the electron-hole interaction, which would suggest that $\Delta_{crystal} = 1/\varepsilon \cdot \Delta_{molecule}$, where $\Delta$ is the exciton binding energy. As before, we take $\varepsilon$ from the RPA calculations – 3.6 for pentacene and 4.0 for PTCDA. $\Delta_{crystal}$ calculated in this way is 0.63 eV and 0.65 eV for pentacene and PTCDA, in very good agreement with our $G_0W_0$/BSE results, as shown in Table 1.

## B. Bandstructure, Densities of States, and Photoemission

It has been often observed, starting with the early work of Hybertsen and Louie,[17] that for low-lying occupied and unoccupied states (separately) in conventional semiconductors, there is an approximately linear relation between the GW QP energies and the starting DFT eigenvalues. Specifically, it is often found that the $G_0W_0$ QP corrections to DFT are roughly k-independent and follow a linear relation, $\varepsilon^{QP} - \varepsilon^{DFT} = s\varepsilon^{DFT} + \delta$. This can be rationalized in terms of a first order correction between DFT eigenvalues and QP excitation energies.[16, 17, 82] The shift-factor $\delta$ opens up the fundamental bandgap, correcting for the well-known DFT bandgap error;[16] the stretch factor, $s$, compensates for the tendency of standard DFT functionals to compress the QP spectra of semiconductors.[69, 83] With a reliable linear relationship determined from a few k-points, the GW-corrected band structure and density of states (DOS) can be obtained without the need for explicit evaluation of GW corrections over a prohibitively large number of $k$-points. We note that empirically shifted and stretched DFT spectra can also mimic GW, and in fact have been used to obtain quantitative agreement with photoemission for organic systems (e.g. Refs. [84, 85]). However, a posteriori shifts are clearly not predictive, and for pentacene and PTCDA, we show below that such shifts cannot address the physical uncertainties regarding the relations between photoemission data and bulk gaps.

For pentacene, we find that a linear relationship between QP corrections and DFT eigenvalues is indeed obtained using either PBE or HSE as the DFT starting point. However, for PTCDA a linear relation is found only when using HSE as a starting point. Because the oxygen-dominated orbitals of PTCDA suffer from a large self-interaction error,[27, 86, 87] its π-type orbitals require a QP correction that is generally smaller than that of the oxygen-dominated orbitals. With HSE, self-interaction errors are, to a large extent, mitigated, and thus using HSE as a starting point restores the approximately linear relationship. For pentacene, self-interaction errors are known to be relatively small,[86, 87] which explains why either starting functional is appropriate. An added benefit of using HSE is that the perturbative $G_0W_0$ correction is quantitatively more accurate if applied to orbitals that do not suffer from excessive delocalization due to self-interaction errors.[88, 89]

For pentacene, with a PBE starting point $\delta$ and $s$ are found to be 0.19 (0.20) and -0.66 eV (0.62), respectively, for the occupied (unoccupied) states, where the zero is taken to be the middle of the gap. The computed band structure for states near the gap then agree very well with that reported previously.[25] However, the stretch factor for the unoccupied states differs from the 0.10 reported in Ref. [25], likely due to the number of states that were considered in that work for the linear fit. With the HSE starting point, $\delta$ and $s$ are found to be 0.01 (0.008) and -0.54 eV (0.56), respectively, for the occupied (unoccupied) states. The smaller values of $\delta$ and $s$ reflect the larger bandgap and bandwidth, respectively, expected with a hybrid functional.[82] For PTCDA with the PBE starting point, $s$ and $\delta$ are found to be 0.28 (0.22) and -0.50 eV (0.56), respectively, for the π-type occupied (unoccupied) states. With an HSE starting point, all low-lying occupied (unoccupied) states can be described by a linear relationship with a value of $s$ and $\delta$ of 0.02 (0.006) and -0.56 (0.57), respectively.

GW-corrected band structures and densities of states (DOS) are obtained by applying the above linear relationship to the DFT-PBE orbital energies (see Fig. S1). The computed bandstructure shows non-negligible dispersion (highest occupied/lowest unoccupied bandwidth of 0.4/0.7eV for pentacene and 0.2/0.2 eV for PTCDA), though the bands are relatively flat when compared with inorganic semiconductors. We note that while the bandgap is slightly affected by the GW starting point, the bandwidth is nearly unchanged (within 0.05 eV) when $G_1W_1$(PBE) or $G_0W_0$(HSE) is applied. Interestingly, the relative magnitude of the bandwidth between the two crystals is consistent with the relative extent of the exciton in each. PTCDA, with a smaller bandwidth, also shows a more localized exciton in Fig. 2. The computed bandwidths for the S-phase of pentacene agree well with previous $G_0W_0$ calculations.[25] Moreover, angle-resolved photoemission experiments on films measure an occupied bandwidth of 0.25 eV for pentacene on metals, at 120K, with an observed increase of the bandwidth as the temperature is lowered[90,91] (possibly due to polaronic effects[92]) and a strong sensitivity to the polymorph measured.[13] The measured HOMO bandwidth is 0.2 eV for PTCDA on $MoS_2$.[93]

Photoemission spectroscopy (PES) is a principal electronic structure characterization tool that, if one can neglect high-order scattering events and final state effects, provides the density of states of a film deposited on a conducting substrate. PES is a surface sensitive technique that probes occupied states by ejecting electrons from the film, typically using ultraviolet light. A complementary spectroscopy is inverse PES (IPES), which probes unoccupied states by injecting electrons into the film, resulting in photon emission. Taken together, PES and IPES then offer a direct comparison to our QP spectra computed with the GW approximation. However, the comparison between the PES/IPES data and the bulk QP DOS is complicated by two additional considerations: first, photoemission typically probes the electronic structure near or at the surface, with electronic structure that may be different than that of the bulk; second, measurements are taken at finite temperature, averaging over a dynamic lattice. Thus, careful analysis of both the measured and predicted spectra is necessary.

Digitized[94] PES spectra extracted from Refs. [7, 8, 10, 12] are shown in Figs. 3.a1 and 3.b1 for pentacene and PTCDA, respectively. The peaks are quite broad, with a full width half maximum (FWHM) > 1.0 eV. In one study the broadening was attributed to disorder in the samples leading to local variations in energy; in this interpretation, the energies of the molecular levels were assumed to have zero dispersion, and the HOMO and LUMO energies could then be defined at the maximum of the peaks in the spectra.[10] In a separate study, the band extrema associated with the molecular orbital energies were defined at the edge onset of the peaks,[11, 12] with the broadening attributed to dynamic polarization effects. Accordingly, extraction of orbital energies from the spectra is controversial, yielding values for the transport gap that vary by over 1 eV[10-12] (see Table 2). In practice, the broadening could arise from the combination of static and dynamic disorder, dispersion, and instrumental resolution.

The ambiguities associated with the broadening lead to additional controversies in the interpretation of the implied transport gaps. In a "peak-to-peak" interpretation, it has been suggested that a reduction in polarization associated with reduced screening at the surface (relative to bulk) shifts occupied and unoccupied states apart with respect to the bulk by 0.5 and 0.6 eV for pentacene and PTCDA, respectively, and that final state vibrational effects shift measured excitation energies to higher binding energy by at most the Franck-Condon factor (estimated to be ~ 0.1 eV per orbital[7]), resulting in a PES/IPES measured gap that is artificially too large by about 0.2 eV; the combination of these effects leads to a reported bulk transport gap of 2.7 and 3.2 eV for pentacene[7] and PTCDA,[10] respectively. On the other hand, a strict "edge-to-edge" interpretation, taking polarization changes at the surface to be negligible[95, 96] and neglecting vibrational effects, leads to a transport gap of 2.2 eV for pentacene (as estimated from the spectra of Ref. [7]) and is reported to be 2.5-2.7 eV for PTCDA.[10-12] Both interpretations lead to transport gaps that are in the range of those extracted from (photo) conductivity measurements[11, 97] but are clearly based on differing and opposing physical intuition. Moreover, our GW calculations of the bulk transport gap, with differing GW start points, yield 2.2-2.4 eV for pentacene and 2.7-3.0 eV for PTCDA, in excellent agreement with both the above-mentioned transport-based data and the experimental spread of photoemission based data.

**Table 2:** Comparison of the edge-to-edge and peak-to-peak gaps, in eV, extracted from the $G_0W_0$(HSE) density of states and photoemission spectra.

|  | Pentacene | PTCDA |
|---|---|---|
| **Photoemission** | | |
| Peak-to-peak gap | 3.4[7] | 3.8-4.1[10-12] |
| Peak-to-peak gap, with surface and temperature corrections | 2.7[7] | 3.2[10] |
| Edge-to-edge gap | 2.2[a] | 2.5-2.7[11, 12] |
| **Theory** | | |
| Peak-to-peak gap (broadened DOS)[b] | 2.8 | 3.2 |
| Edge-to-edge gap (broadened DOS)[b] | 1.7 | 1.9 |
| Predicted transport gap | 2.4 | 3.0 |

[a] Estimated from photoemission spectrum of Ref. [7].
[b] DOS broadened by convolution with a Gaussian to account for experimental conditions. See text for details.

Direct comparison of our broadened $G_0W_0$(HSE) DOS (Figs. 3.a3 and 3.b3) with published PES and IPES data uncovers intriguing insight regarding the experimental uncertainties. For pentacene, the calculated spectrum is convolved with a Gaussian, whose width is estimated based on 0.25 eV broadening due to temperature effects[98] and 0.2 (0.4) eV broadening due to instrumental resolution for occupied (unoccupied) states. Although vibronic progression can lead to an asymmetric broadening of the photoemission spectrum,[13] a uniform Gaussian broadening leads to very good agreement with experimental peak widths, indicating that a combination of dispersion, finite temperature broadening, and instrumental resolution can explain the peak widths here. For PTCDA, the occupied states had to be broadened by 0.6 eV to obtain agreement with experiment, possibly attributable to additional disorder in the PTCDA crystal. Remarkably, if the photoemission occupied and unoccupied states are rigidly shifted together by 0.6 eV for pentacene and, for comparison with Ref. [10], 0.9 eV for PTCDA (Figs. 3.a2 and 3.b2), the GW and PES spectra are in excellent agreement over an energy range of greater than 4 eV on either side of the gap, with the peak maxima deviating by ~ 0.1 eV, and at most 0.2 eV for the higher-lying unoccupied states (which are subject to more experimental uncertainty). The need for a rigid "scissors" shift reflects the fact that the calculated peak-to-peak gaps, taken from the broadened DOS, (also presented in Table 2) are smaller than experiment by 0.6 (0.6-0.9) eV for pentacene (PTCDA). This rigid shift is consistent with the hypothesis of Hill et al.[10] Furthermore, the level of agreement between theory and experiment upon this shift suggests that it must have a physical origin, which we now turn to investigating.

As mentioned previously, the fact that photoemission is surface sensitive points to a possible origin of this discrepancy, namely that the gap is simply larger – and closer to its gas-phase value – for states near the surface due to incomplete screening from the bulk. To investigate this possibility with GW calculations, we consider a model pentacene surface,[99] inspired by experimental geometries.[100-102] We compute GW corrections to the DFT electronic structure for surface layers consisting of two different phases of pentacene in the thick film orientation: the bulk triclinic phase with two inequivalent molecules per unit cell (T2) and the cubic phase with one molecule per unit cell (C1). All phases have the same inter-molecular spacing, equivalent to bulk. We study T2 as a 1-layer surface, with vacuum on both sides, while both 1-layer and 2-layer C1 are considered. The 1-layer calculation provides an upper bound on the difference between surface and bulk polarization. For the 1-layer surface, the transport gap is larger than that of the bulk by 0.2-0.4 eV for the different phases, while the gap of 2-layer C1 is that of the bulk crystal. The model geometries used here, with the limited size that can be feasibly studied with GW, provide qualitative, rather than quantitative, insight into the significance of the incomplete screening at the surface, but cumulatively indicate that this effect is not very large on the fundamental gap. The gap at the surface is at most 0.4 eV larger than that of the bulk, and practically less than that since measurements are taken on a film (and not an isolated 1-layer slab).

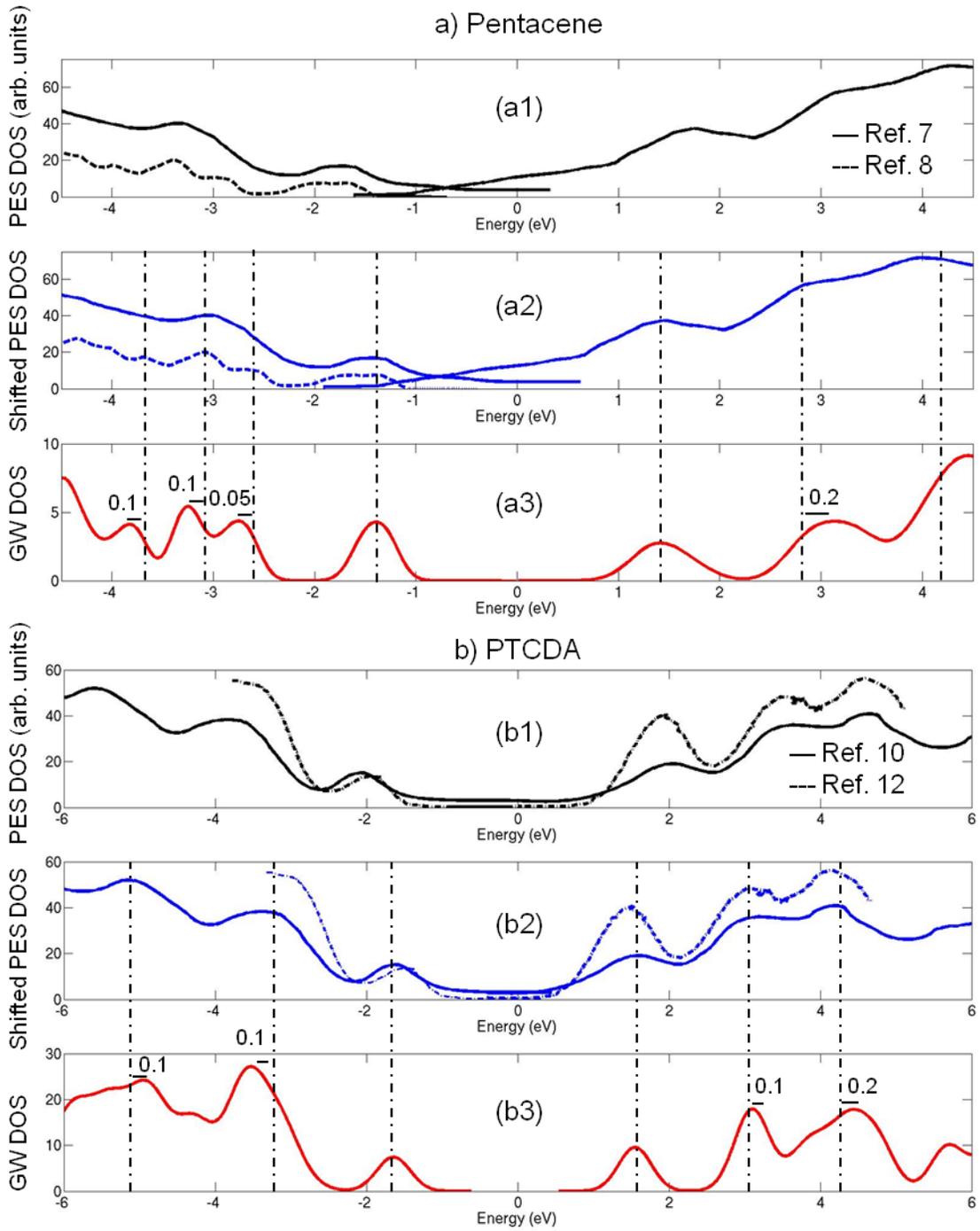

**Figure 3**: Comparison of the GW broadened density of states (a3 and b3) with (inverse) photoemission (PES) data extracted from Refs. [7, 8] (a1) and [10, 12] (b1), for pentacene and PTCDA, respectively. The spectra are aligned such that zero is at the center of the gap. For Ref. [8], because unoccupied states are unavailable, the spectrum is shifted so that the highest occupied peak lines up with Ref. [7]. The calculated

and experimental spectra agree very well when the photoemission gap is rigidly shifted by 0.6 eV for (a2) pentacene and 0.9 eV for (b2) PTCDA.

In addition, we estimate the difference in polarization energy between surface and bulk in a different way, from numerical solution of Poisson's equation of a sphere of radius R near a planar boundary between a dielectric and vacuum (COMSOL Multiphysics Package[103]), with the computational details given in the SI. The difference in electrostatic energy when the sphere is within the bulk and when it is at the surface is 0.1 eV, indicating that the gap is 0.2 eV larger at the surface than within the bulk. This value is smaller than the 0.5 eV estimated in Ref.[7], but in good agreement with the charge redistribution model of Soos and coworkers,[76] and the above upper bound obtained from the GW calculations. For PTCDA, electrostatics again predicts a 0.2 eV difference between the surface and bulk gaps. However, the charge redistribution model predicts a larger effect, with the gap at the surface being 0.4 eV larger than that within the bulk.[75]

In summary, we can expect that occupied and unoccupied states at the surface will be shifted apart by, at most, 0.1-0.2 eV due to reduced polarization, resulting in a larger gap than in bulk by twice that amount (i.e. up to 0.2-0.4 eV). Thus, surface effects can explain a fraction of the rigid shift necessary to bring experimental and theoretical peak-to-peak gaps in agreement. The remainder of the discrepancy between theory and experiment must be due to a combination of additive effects at the 0.1 eV level, including bandwidth narrowing (~0.1-0.2eV[65]) final state vibrational effects that are estimated to open up the gap by ~0.2 eV,[7, 10] and the fact that a static 0K calculation is compared to finite-temperature measurements (~0.1-0.2eV[104]). Additionally, one should consider that sample-related uncertainties lead to a 0.3 eV spread in the measured peak-to-peak gaps,[10-12] and possible residual inaccuracies of our GW calculations (of order 0.1 eV).

Lastly, we note that although GW does predict the need for a shift of the PES spectrum, a consequence of our predictions for the transport gap and small surface effects is that the peak-to-peak transport gaps (with surface corrections taken from electrostatics and temperature corrections taken as 0.2 eV) are too large when compared with theory, by 0.2-0.4 eV. This is not surprising considering that within the peak-to-peak interpretation, band dispersion, which we predict to be a few tenths of eVs for pentacene and PTCDA, is not accounted for. Though the edge-to-edge interpretation results in transport gaps closer to our calculations and conductivity measurements, it incorporates broadening of the peaks due to finite temperature and instrumental resolution into the gap, and neglects surface effects all together. The relationship between the transport gap and data extracted from PES/IPES is complicated by the convolution of many uncontrolled measurement conditions. More information may be found with low temperature studies that minimize temperature effects, along with further theoretical insight into the magnitude of surface effects and dynamic polarization effects. Our calculations shed new light on a long-standing controversy in the interpretation of photoemission data and indicate that quantitative bulk transport gaps are not obtained directly from surface-sensitive photoemission/inverse photoemission gaps, calling for experiments at higher resolution for further quantification of the detailed role of each of these physical mechanisms.

## IV. Conclusions

In conclusion, we determined the electronic structure and spectroscopic properties of two prototypical organic systems, pentacene and PTCDA, from first principles. This was achieved by combining density functional theory (with the PBE and the HSE functionals), and many-body perturbation theory within the GW approximation and the Bethe-Salpeter equation approach for charged and neutral excitations, respectively. This provided for a detailed and quantitative comparison with photoemission, inverse photoemission, and optical spectroscopy experiments in these systems. Comparing the fundamental and the optical gaps of the gas phase molecules with those of the molecular solid, we found that the differences can be rationalized via polarization effects that can be explained with a simple electrostatic expression. We further found that the solid-state exciton binding energy is significant in both systems, of the order of ~ 0.5 eV. Furthermore, in PTCDA the solid-state exciton is qualitatively similar to that of the gas-phase molecule, whereas for pentacene it is delocalized over several molecules and exhibits charge transfer character. Through detailed band structure and density of states calculations we addressed the interpretation of the photoemission and inverse photoemission data, and the uncertainty in determining the transport gap from these data. Our GW transport gaps in the bulk crystalline phase were 2.4 eV and 3.0 eV for pentacene and PTCDA, respectively, in good agreement with values extrapolated from a variety of experiments. We found that the bulk density of states, obtained from a DFT-based DOS that was shifted and stretched according to the GW results, was in excellent agreement with the photoemission spectrum for a wide energy range of over 4 eV both above and below the fundamental gap, but only if the experimental photoemission/inverse photoemission gap is rigidly closed by 0.6 eV for pentacene and 0.6-0.9 eV for the different published data on PTCDA. We attributed the origin of this rigid shift to a combination of several physical effects, including surface polarization, vibrational contributions, and a dynamical lattice, and to some extent also to residual errors of both theory and experiment. Our results shed new light on a long-standing controversy in the interpretation of photoemission data and calls for experiments at higher resolution for further quantification the detailed role of each of these physical mechanisms.

**Acknowledgements**

Portions of this work were performed at the Molecular Foundry, supported by the Office of Science, Office of Basic Energy Sciences, of the U.S. Department of Energy under Contract No. DE-AC02-05CH11231. We also acknowledge funding from the National Science Foundation through the Network for Computational Nanotechnology (NCN), the Israel Science Foundation, and the United States-Israel Binational Science Foundation (BSF), and acknowledge the National Energy Research Scientific Computing (NERSC) center for computational resources. We are grateful for useful discussions with Antoine Kahn (Princeton University), Achim Schöll (Universität Würzburg), Lothar Weinhardt (Universität Würzburg), and Clemens Heske (University of Nevada at Las Vegas) regarding the interpretation of photoemission spectra, with Sohrab Ismail-Beigi (Yale University) regarding GW calculations for slab configurations, Jack Deslippe, Georgy Samsonidze, and Manish Jain (UC Berkeley) for BerkeleyGW support, and Peter Doak, Pierre Darancet, and Isaac Tamblyn (Molecular Foundry).

**Appendix: Electrostatic calculations of a charged sphere near a dielectric surface**

Here, we model the charged molecule as a hollow sphere of radius R, situated a distance, $d$, from the surface of a linear dielectric (see Fig. A1), with the dielectric constant, $\varepsilon$, set to the RPA value of 3.6 for pentacene and 4.0 for PTCDA. R is determined from the volume per molecule in the crystal (R = $(V*3/4\pi)^{1/3}$) and is 4.36 Å for pentacene and 4.50 Å for PTCDA. The dielectric is represented by a cube of side length 1 $\mu$m. This box size is necessary for the potential at the edges of the box to be zero without the presence of a dielectric. Maxwell's equations are solved numerically on a non-uniform grid, with the grid density chosen such that the polarization energy changes by less than 0.01 eV with increase of grid size.[103]

The polarization energy for a given $d$ is computed as P = $1/2[U(+\infty) - U(-d)]$, where U is the potential induced by the charge. The difference between the bulk and surface electrostatic energies is computed as the difference between P calculated with $|d| \gg |R|$ and $|d| = |R|$.

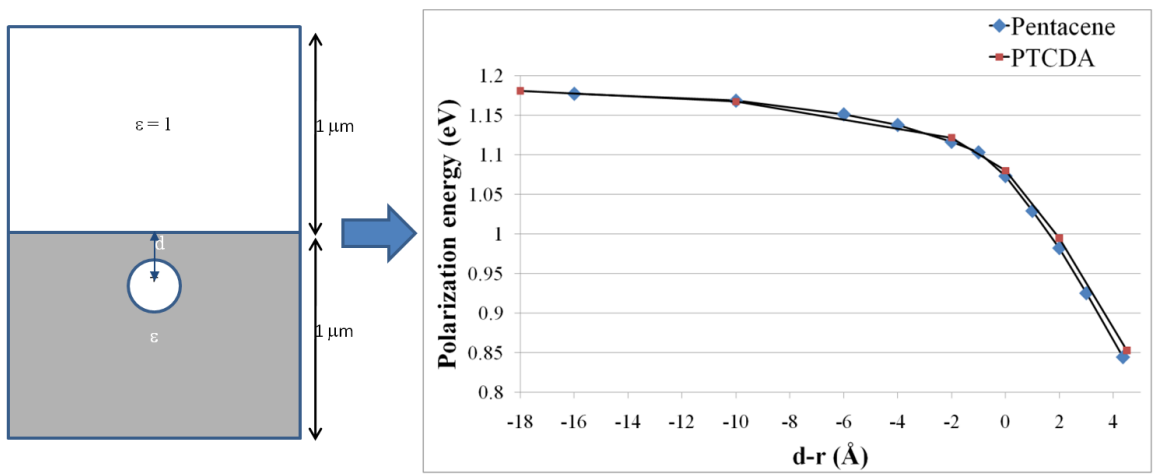

**Figure A1**: The electrostatic energy to charge a sphere within a linear dielectric. Note: the 3D system is projected onto a 2D picture, which is not depicted to scale.

**Supporting Information**

Figure S1 shows the $G_0W_0$(PBE) bandstructure and density of states for the $\pi$-type states of pentacene and PTCDA. The bandstructure and density of states are computed within DFT-PBE and the $G_0W_0$(PBE) shift- and stretch-parameters are applied to the eigenvalues. The band-structures show moderate dispersion for the $\pi$-type states.

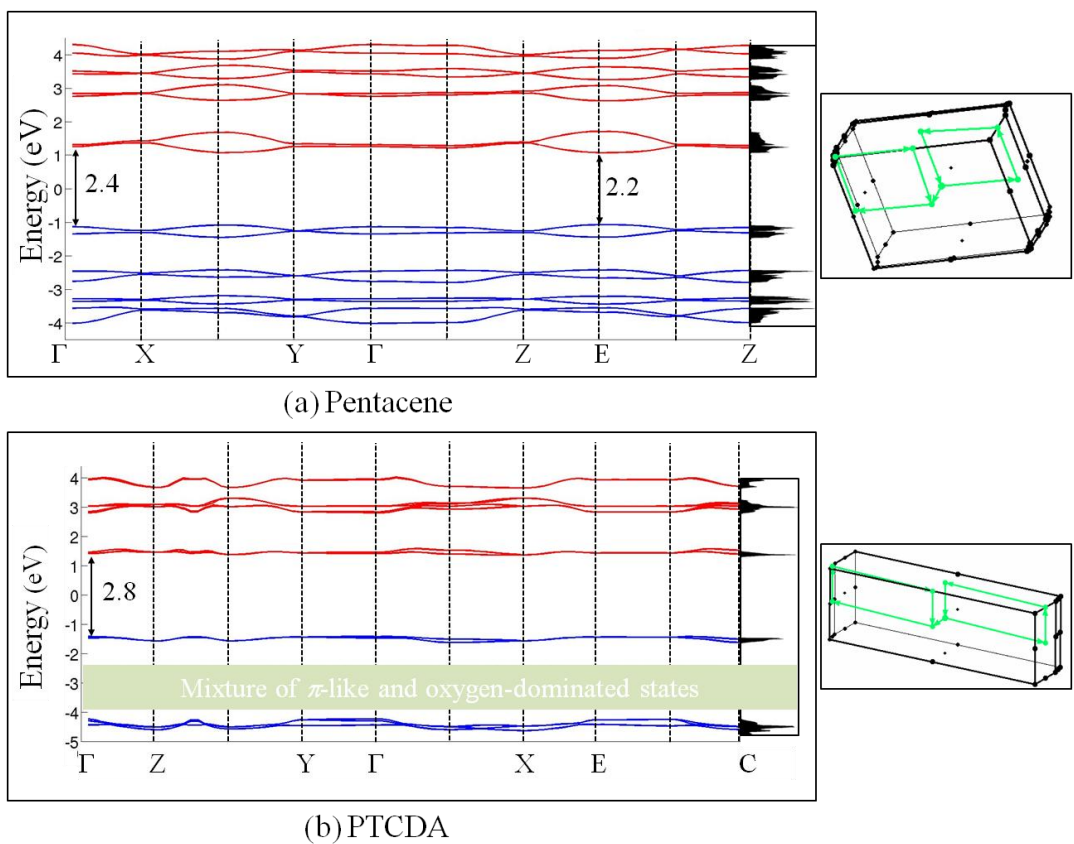

**Figure S1:** $G_0W_0$(PBE) bandstructure and density of states for a) pentacene and b) PTCDA, along with the crystals' Brillouin zone. The energies are aligned such that zero is at mid-gap. For the bandstructure, the occupied and unoccupied states are colored in blue and red, respectively.

# References

[1] S. Forrest, Nature **428**, 911 (2004).
[2] Z. V. Vardeny, A. J. Heeger, and A. Dodabalapur, Synthetic Metals **148**, 1 (2005).
[3] B. Kippelen and J.-L. Bredas, Energy & Environmental Science **2**, 251 (2009).
[4] S. Sanvito, Nat Mater **6**, 803 (2007).
[5] J. A. Rogers, T. Someya, and Y. Huang, Science **327**, 1603 (2010).
[6] G. Witte and G. Wöll, Journal of Materials Research **19**, 1889 (2004).
[7] F. Amy, C. Chan, and A. Kahn, Organic Electronics **6**, 85 (2005).
[8] H. Fukagawa, H. Yamane, T. Kataoka, S. Kera, M. Nakamura, K. Kudo, and N. Ueno, Physical Review B **73**, 245310 (2006).
[9] N. J. Watkins, S. Zorba, and Y. Gao, Journal of Applied Physics **96**, 425 (2004).
[10] I. G. Hill, A. Kahn, Z. G. Soos, and J. R. A. Pascal, Chemical Physics Letters **327**, 181 (2000).
[11] D. R. T. Zahn, G. N. Gavrila, and M. Gorgoi, Chemical Physics **325**, 99 (2006).
[12] S. Krause and et al., New Journal of Physics **10**, 085001 (2008).
[13] N. Ueno and S. Kera, Progress in Surface Science **83**, 490 (2008).
[14] W. E. Pickett, Computer Physics reports **9**, 115 (1989).
[15] J. R. Chelikowsky and M. L. Cohen, *Handbook on Semiconductors* (Elsevier, Amsterdam, 1992).
[16] S. Kümmel and L. Kronik, Reviews of Modern Physics **80**, 3 (2008).
[17] M. S. Hybertsen and S. G. Louie, Physical Review B **34**, 5390 (1986).
[18] M. Rohlfing and S. G. Louie, Physical Review B **62**, 4927 (2000).
[19] M. S. Hybertsen and S. G. Louie, Physical Review B **32**, 7005 (1985).
[20] S. G. Louie, in *Topics in Computational Materials Science*, edited by C. Y. Fong (World Scientific, Singapore, 1997), p. 96.
[21] W. G. Aulbur, L. Jönsson, and J. W. Wilkins, in *Solid State Physics*, edited by H. Ehrenreich and F. Spaepen (Academic Press, 1999), Vol. 54, p. 1.
[22] P. García-González and R. W. Godby, Computer Physics Communications **137**, 108 (2001).
[23] E. L. Shirley and S. G. Louie, Physical Review Letters **71**, 133 (1993).
[24] J. W. van der Horst, P. A. Bobbert, M. A. J. Michels, G. Brocks, and P. J. Kelly, Physical Review Letters **83**, 4413 (1999).
[25] M. L. Tiago, J. E. Northrup, and S. G. Louie, Physical Review B **67**, 115212 (2003).
[26] N. Sai, M. L. Tiago, J. R. Chelikowsky, and F. A. Reboredo, Physical Review B **77**, 161306 (2008).
[27] N. Dori, M. Menon, L. Kilian, M. Sokolowski, L. Kronik, and E. Umbach, Physical Review B (Condensed Matter and Materials Physics) **73**, 195208 (2006).
[28] M. L. Tiago, P. R. C. Kent, R. Q. Hood, and F. A. Reboredo, The Journal of Chemical Physics **129**, 084311 (2008).
[29] M. Palummo, C. Hogan, F. Sottile, P. Bagala, and A. Rubio, The Journal of Chemical Physics **131**, 084102 (2009).
[30] M. L. Tiago and F. A. Reboredo, Physical Review B **79**, 195410 (2009).
[31] D. Rocca, D. Lu, and G. Galli, The Journal of Chemical Physics **133**, 164109 (2010).

32 X. Blase, C. Attaccalite, and V. Olevano, Physical Review B **83**, 115103 (2011).
33 S.-H. Ke, Physical Review B **84**, 205415 (2011).
34 R. G. Endres, C. Y. Fong, L. H. Yang, G. Witte, and C. Wöll, Computational Materials Science **29**, 362 (2004).
35 K. Hummer and C. Ambrosch-Draxl, Physical Review B **71**, 081202 (2005).
36 E. Engel, K. Schmidt, D. Beljonne, J.-L. Bredas, J. Assa, H. Frob, K. Leo, and M. Hoffmann, Physical Review B **73**, 245216 (2006).
37 E. S. Kadantsev, M. J. Stott, and A. Rubio, The Journal of Chemical Physics **124**, 134901 (2006).
38 P. Sony and A. Shukla, Physical Review B **75**, 155208 (2007).
39 C. Ambrosch-Draxl, D. Nabok, P. Puschnig, and C. Meisenbichler, New Journal of Physics **11**, 125010 (2009).
40 G. Cappellini, G. Malloci, and G. Mulas, Superlattices and Microstructures **46**, 14 (2009).
41 B. Hajgato, D. Szieberth, P. Geerlings, F. De Proft, and M. S. Deleuze, The Journal of Chemical Physics **131**, 224321 (2009).
42 L. Huang, D. Rocca, S. Baroni, K. E. Gubbins, and M. B. Nardelli, The Journal of Chemical Physics **130**, 194701 (2009).
43 R. Mondal, C. Tonshoff, D. Khon, D. C. Neckers, and H. F. Bettinger, Journal of the American Chemical Society **131**, 14281 (2009).
44 P. K. Nayak and N. Periasamy, Organic Electronics **10**, 1396 (2009).
45 Y. Yi, V. Coropceanu, and J.-L. Bredas, Journal of the American Chemical Society **131**, 15777 (2009).
46 L. Hedin, Physical Review **139**, A796 (1965).
47 M. S. Hybertsen and S. G. Louie, Physical Review B **35**, 5585 (1987).
48 F. Fuchs, J. Furthmüller, F. Bechstedt, M. Shishkin, and G. Kresse, Physical Review B **76**, 115109 (2007).
49 M. Shishkin and G. Kresse, Physical Review B **74**, 035101 (2006).
50 M. Shishkin, M. Marsman, and G. Kresse, Physical Review Letters **99**, 246403 (2007).
51 J. P. Perdew, K. Burke, and M. Ernzerhof, Physical Review Letters **77**, 3865 (1996).
52 J. Heyd, G. Scuseria, E. , and M. Ernzerhof, The Journal of Chemical Physics **118**, 8207 (2003).
53 J. Heyd, G. Scuseria, E. , and M. Ernzerhof, The Journal of Chemical Physics **124**, 219906 (2006).
54 A. V. Krukau, O. A. Vydrov, A. F. Izmaylov, and G. E. Scuseria, The Journal of Chemical Physics **125**, 224106 (2006).
55 S. Theo, K. Christian, H. S. Jan, B. Bertram, C. H. Robert, B. Steffen, and A. T. Gordon, Angewandte Chemie International Edition **40**, 1732 (2001).
56 K. Tojo and J. Mizuguchi, Z.Kristallogr.-New Cryst.Struct. **217**, 45 (2002).
57 R. B. Campbell, J. M. Robertson, and J. Trotter, in *Acta Crystallographica*, 1962), Vol. 15, p. 289.
58 D. Nabok, P. Puschnig, and C. Ambrosch-Draxl, Physical Review B **77**, 245316 (2008).
59 N. Marom, A. Tkatchenko, M. Scheffler, and L. Kronik, Journal of Chemical Theory and Computation **6**, 81 (2009).
60 A. Tkatchenko, L. Romaner, O. T. Hofmann, E. Zojer, C. Ambrosch-Draxl, and M. Scheffler, MRS Bulletin **35**, 435 (2010).
61 G. Kresse and J. Furthmüller, Physical Review B **54**, 11169 (1996).

| | |
|---|---|
| 62 | G. Kresse and J. Furthmüller, Computational Materials Science **6**, 15 (1996). |
| 63 | G. Kresse and D. Joubert, Physical Review B **59**, 1758 (1999). |
| 64 | J. Deslippe, G. Samsonidze, D. A. Strubbe, M. Jain, M. Cohen, L., and S. G. Louie, Computer Physics Communications **in press** (2011). |
| 65 | G. Paolo and et al., Journal of Physics: Condensed Matter **21**, 395502 (2009). |
| 66 | N. Troullier and J. L. Martins, Physical Review B **43**, 1993 (1991). |
| 67 | N. Sato, H. Inokuchi, and E. A. Silinsh, Chemical Physics **115**, 269 (1987). |
| 68 | L. Crocker, T. Wang, and P. Kebarle, Journal of the American Chemical Society **115**, 7818 (1993). |
| 69 | R. O. Jones and O. Gunnarsson, Reviews of Modern Physics **61**, 689 (1989). |
| 70 | T. Stein, H. Eisenberg, L. Kronik, and R. Baer, Physical Review Letters **105**, 266802 (2010). |
| 71 | S. Refaely-Abramson, R. Baer, and L. Kronik, Physical Review B **84**, 075144 (2011). |
| 72 | J. B. Neaton, M. S. Hybertsen, and S. G. Louie, Physical Review Letters **97**, 216405 (2006). |
| 73 | N. Sato, K. Seki, and H. Inokuchi, Journal of the Chemical Society, Faraday Transactions 2: Molecular and Chemical Physics **77**, 1621 (1981). |
| 74 | Z. G. Soos, E. V. Tsiper, and R. A. Pascal, Chemical Physics Letters **342**, 652 (2001). |
| 75 | E. V. Tsiper, Z. G. Soos, W. Gao, and A. Kahn, Chemical Physics Letters **360**, 47 (2002). |
| 76 | E. V. Tsiper and Z. G. Soos, Physical Review B **68**, 085301 (2003). |
| 77 | J. E. Norton and J.-L. Bredas, Journal of the American Chemical Society **130**, 12377 (2008). |
| 78 | To find a high probability location for the hole, we integrate out the electronic degrees of freedom of the two-particle wavefunction for various fixed hole positions in the supercell. We sampled hole positions within one lattice vector of the center of the 4x4x2 (2x3x2) supercell for pentacene (PTCDA), with a uniform spacing of 1-3 Å. |
| 79 | L. Sebastian, G. Weiser, and H. Bässler, Chemical Physics **61**, 125 (1981). |
| 80 | S. Haas, H. Matsui, and T. Hasegawa, Physical Review B **82**, 161301 (2010). |
| 81 | M. Hoffmann, K. Schmidt, T. Fritz, T. Hasche, V. M. Agranovich, and K. Leo, Chemical Physics **258**, 73 (2000). |
| 82 | T. Körzdörfer and S. Kümmel, Physical Review B **82**, 155206 (2010). |
| 83 | N. Marom and L. Kronik, Applied Physics A: Materials Science & Processing **95**, 159 (2009). |
| 84 | J. Hwang, E.-G. Kim, J. Liu, J.-L. Bredas, A. Duggal, and A. Kahn, The Journal of Physical Chemistry C **111**, 1378 (2006). |
| 85 | L. Segev, A. Salomon, A. Natan, D. Cahen, L. Kronik, F. Amy, C. K. Chan, and A. Kahn, Physical Review B **74**, 165323 (2006). |
| 86 | T. Körzdörfer, S. Kümmel, N. Marom, and L. Kronik, Physical Review B **79**, 201205 (2009). |
| 87 | T. Körzdörfer, S. Kümmel, N. Marom, and L. Kronik, Physical Review B **82**, 129903(E) (2010). |
| 88 | N. Marom, X. Ren, J. E. Moussa, J. R. Chelikowsky, and L. Kronik, Physical Review B **84**, 195143 (2011). |
| 89 | P. Rinke, A. Qteish, J. Neugebauer, C. Freysoldt, and M. Scheffler, New Journal of Physics **7**, 126 (2005). |

90   N. Koch, A. Vollmer, I. Salzmann, B. Nickel, H. Weiss, and J. P. Rabe, Physical Review Letters **96**, 156803 (2006).
91   S. Berkebile, G. Koller, A. J. Fleming, P. Puschnig, C. Ambrosch-Draxl, K. Emtsev, T. Seyller, J. Riley, and M. G. Ramsey, Journal of Electron Spectroscopy and Related Phenomena **174**, 22 (2009).
92   K. Hannewald, V. M. Stojanovic, J. M. T. Schellekens, P. A. Bobbert, G. Kresse, and J. Hafner, Physical Review B **69**, 075211 (2004).
93   H. Yamane, S. Kera, K. K. Okudaira, D. Yoshimura, K. Seki, and N. Ueno, Physical Review B **68**, 033102 (2003).
94   B. Tummers, DataThief III., http://datathief.org (2006).
95   M. B. Casu, Y. Zou, S. Kera, D. Batchelor, T. Schmidt, and E. Umbach, Physical Review B **76**, 193311 (2007).
96   M. B. Casu, physica status solidi (RRL) – Rapid Research Letters **2**, 40 (2008).
97   E. A. Silinsh, V. A. Kolesnikov, I. J. Muzikante, and D. R. Balode, physica status solidi (b) **113**, 379 (1982).
98   Ab initio Molecular Dynamics calculations on the isolated pentacene and PTCDA molecule show a spread of 0.5 eV in the gap at finite temperature, due to vibrational motion of the molecule.
99   The photoemission spectrum plotted in Fig. 3 is taken on thick pentacene films on Au, which grow in the polycrystalline phase, with the long molecular axis perpendicular to the surface.
100  H. Ozaki, The Journal of Chemical Physics **113**, 6361 (2000).
101  P. G. Schroeder, C. B. France, J. B. Park, and B. A. Parkinson, The Journal of Physical Chemistry B **107**, 2253 (2003).
102  G. Beernink, T. Strunskus, G. Witte, and W. Ch,  (AIP, 2004), Vol. 85, p. 398.
103  http://www.comsol.com/products/multiphysics.
104  Within G0W0(PBE), a +/-2% increase in lattice constant, which may reasonably occur at finite temperatures, changes the gap by +/-0.15 eV.
105  T. M. Halasinski, D. M. Hudgins, F. Salama, L. J. Allamandola, and T. Bally, The Journal of Physical Chemistry A **104**, 7484 (2000).
106  S. P. Park, S. S. Kim, J. H. Kim, C. N. Whang, and S. Im, Applied Physics Letters **80**, 2872 (2002).
107  D. Faltermeier, B. Gompf, M. Dressel, A. K. Tripathi, and J. Pflaum, Physical Review B **74**, 125416 (2006).
108  T. Jentzsch, H. J. Juepner, K. W. Brzezinka, and A. Lau, Thin Solid Films **315**, 273 (1998).
109  M. Wewer and F. Stienkemeier, Physical Review B **67**, 125201 (2003).
110  V. Bulovic, P. E. Burrows, S. R. Forrest, J. A. Cronin, and M. E. Thompson, Chemical Physics **210**, 1 (1996).
111  M. Leonhardt, O. Mager, and H. Port, Chemical Physics Letters **313**, 24 (1999).